\documentclass[a4paper]{jpconf}
\usepackage{graphicx}
\begin{document}
\title{A stochastic method of solution of the  Parker transport equation}

\author{A. Wawrzynczak$^{1}$, R. Modzelewska$^{2}$, A. Gil$^{2}$}

\address{$^{1}$Institute of Computer Sciences, Siedlce University, Poland, \\ $^{2}$Institute of Mathematics and Physics, Siedlce University, Poland.}

\ead{awawrzynczak@uph.edu.pl, renatam@uph.edu.pl, gila@uph.edu.pl,}

\begin{abstract}
We present the stochastic model of the galactic cosmic ray (GCR) particles transport in the heliosphere. Based on the solution of the Parker transport equation  we developed models of the short-time variation of the GCR intensity, i.e. the Forbush decrease (Fd) and the 27-day variation of the GCR intensity.
Parker transport equation being the Fokker-Planck type equation delineates non-stationary transport of charged particles in the turbulent medium. The presented approach of the numerical solution is grounded on solving of the set of equivalent stochastic differential equations (SDEs). We demonstrate the method of deriving from Parker transport equation the corresponding SDEs in the heliocentric spherical coordinate system for the backward approach. Features indicative the preeminence of the backward approach over the forward is stressed. We compare the outcomes of the stochastic model of the Fd and 27-day variation of the GCR intensity with our former models established by the finite difference method. Both models are in an agreement with the experimental data.

\end{abstract}

\section{Introduction}
First stochastic equation (Langevin's) was linked with the Newton's principle \cite{La}. From the beginning of the XX century, the stochastic approach became more useful for describing physical random processes. The stochastic differential equations (SDEs) in conjunction with various Monte Carlo technics are broadly used in many fields, like physics, finance, biology, chemistry, engineering, or management science. The main statistical characteristic is the representation of the solution of the Fokker-Planck type equation as a probability density distributions. Employment of probabilistic description with Monte Carlo simulations allow to reduce the solution of the partial differential equation (PDE) describing the analyzed phenomena to the  integration of SDEs.\\ 
 We apply stochastic methodology to model the galactic cosmic rays (GCR) transport in the heliosphere. Algorithms used in the particle transport simulations are mainly based on finite difference methods (e.g.~\cite{KJ}). Galerkin methods are used for solving time-dependent high order PDEs (e.g.~\cite{YS}). The homotopy perturbation method \cite{He} was recently developed for the numerical solution of various linear and nonlinear PDEs. Also exists approach to the PDE solution grounded on particle methods, characterized by low numerical dispersion \cite{CKM}.\\
During the propagation through the heliosphere, GCR particles are modulated by the solar wind and heliospheric magnetic field (HMF). Modulation of the GCR is a result of action of four primary processes: convection by the solar wind, diffusion on irregularities of HMF, particles drifts in the non-uniform magnetic field  and adiabatic cooling. Transport of the GCR particles in the heliosphere can be described by the Parker transport equation  \cite{Pa}:
\begin{eqnarray} \label{ParkerEq}
\frac{\partial f}{\partial t}=\vec{\nabla}\cdot (K_{ij} ^{S}\cdot \vec{\nabla}f)-(\vec{v}_{d}+\vec{U})\cdot \vec{\nabla} f+\frac{R}{3}(\vec{\nabla} \cdot \vec{U})\frac{\partial f}{\partial R},
\end{eqnarray}
where $f=f(\vec{r}, R, t) $ is an omnidirectional distribution function of three spatial coordinates $\vec{r}=(r,\theta,\varphi)$, particles rigidity $R$ and time $t$;  $\vec{U}$  is solar wind velocity, $\vec{v}_{d}$  the drift velocity, and $K_{ij} ^{S}$ is the symmetric part of the diffusion tensor of the GCR particles.\\
Employing the stochastic approach to solving the Parker transport equation is not the latest idea (\cite{Zhang1999},\cite{Gerv}). However, majority of models presented in the literature are used to determine the simulated spectra and compare it with experimental observations carried out by space probes as, Voyagers, AMS, BESS and PAMELA (e.g.~\cite{Bobik}-\cite{Guo}).
In this paper, we present models in which we not only reproduce the proton spectra  but also simulate the short time GCR variations i.e. the Forbush decrease (Fd) and 27-day variation. Additionally we compare the results of the stochastic modeling with our previous well grounded models developed by using finite difference method (FDM) of solution the Parker transport equation (e.g.~\cite{WA08}-\cite{Gil14}).
To perform the reliable comparison between two models we consider the same coefficients and parameters in the baseline Parker transport equation. Furthermore, the parameters included attaining the GCR variations from the model are obtained by approximation of the experimental data. \\
The aim of our paper is twofold. The first is to compose a consistent mathematical model of the GCR transport in the heliosphere by means of the SDEs. The second is to employ the created model to simulate the short-term variations being in agreement with the experimental data.

\begin{figure}[!h]
  \begin{center}
\includegraphics[width=0.6\hsize]{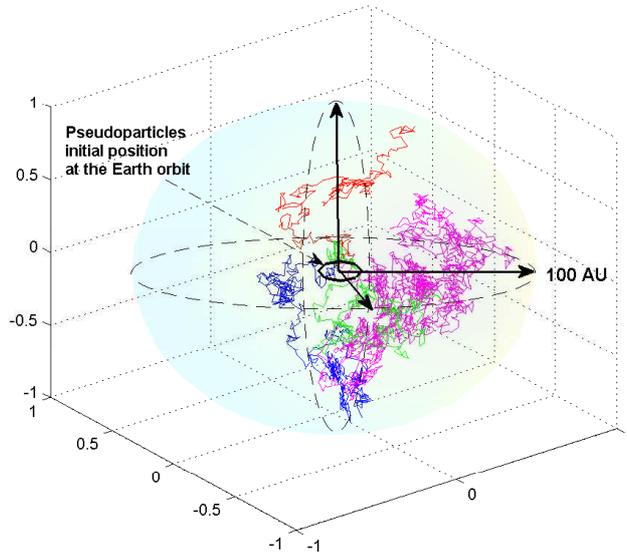}
\end{center}
\caption{\label{fig:fig1} The sample pseudoparticles trajectories within the heliosphere.}
\end{figure}

\begin{figure}[!h]
  \begin{center}
\includegraphics[width=1\hsize]{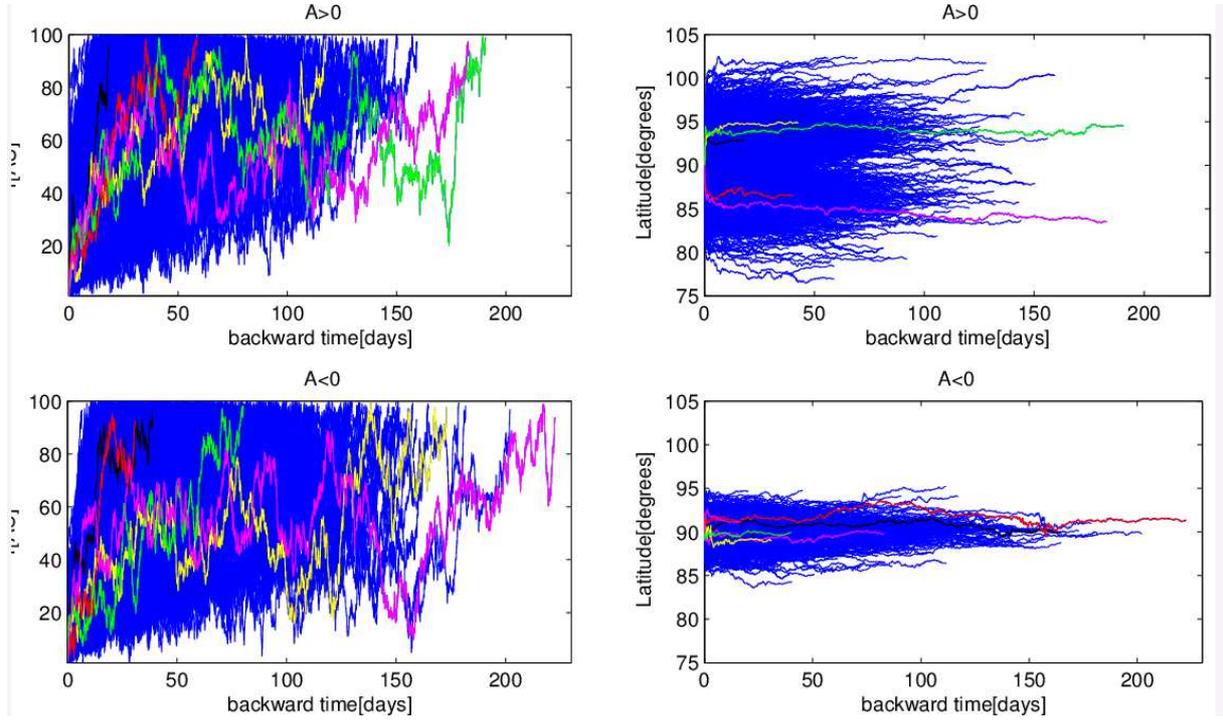}
\end{center}
\caption{\label{fig:fig2} Trajectories of the pseudoparticles initialized with rigidity 10 GV from position $r=1AU$, $\theta=90^{\circ}$, $\varphi=180^{\circ}$ for A$>$0 and A$<$0 solar magnetic cycle.}
\end{figure}

\begin{figure}[!h]
  \begin{center}
\includegraphics[width=0.95\hsize]{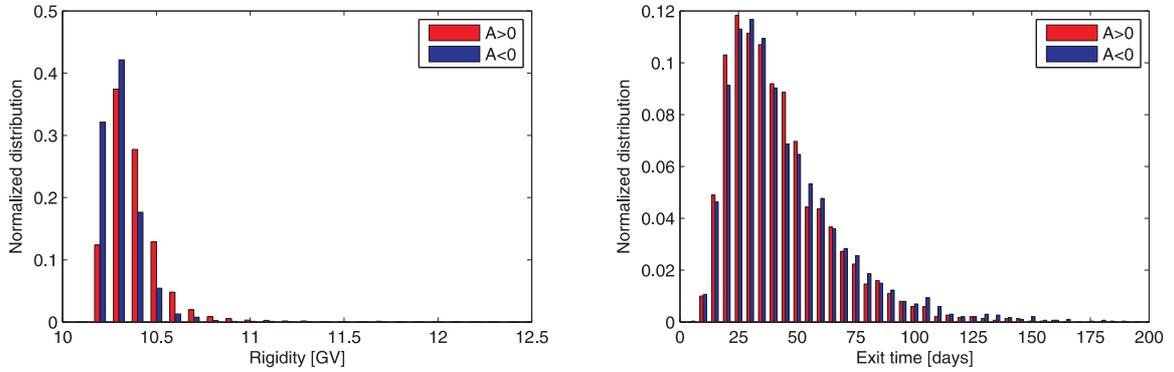}
\end{center}
\caption{\label{fig:fig3} The histograms for the particles rigidity and exit time for the pseudoparticles initialized with rigidity 10 GV from position $r=1AU$, $\theta=90^{\circ}$, $\varphi=180^{\circ}$ for A$>$0 and A$<$0 solar magnetic cycle.}
\end{figure}

\begin{figure}[!h]
  \begin{center}
\includegraphics[width=0.95\hsize]{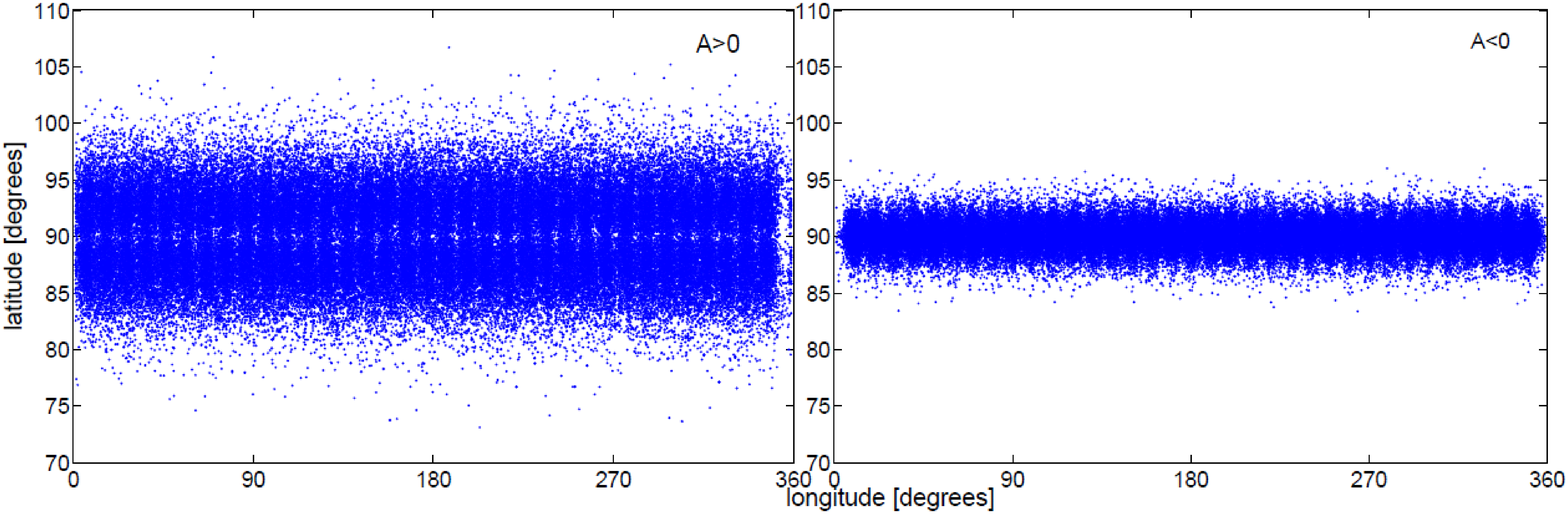}
\end{center}
\caption{\label{fig:scatter} Latitude vs. longitude distribution of simulated pseudoparticles (protons) for $A>0$ and $A<0$ solar magnetic cycle.}
\end{figure}

\section{Stochastic approach}
In order to model the GCR transport in the heliosphere using stochastic methods equivalent SDEs must be obtained. In the first step, Parker transport equation (Eq. \ref{ParkerEq}) should be converted to a general form of Fokker-Planck equation (FPE). Depending on the direction of integration FPE, can be expressed in two forms \cite{Ga}:\\
time-backward:
\small
\begin{eqnarray}
\label{backwardFPE}
\frac{\partial F}{\partial t}=\sum _{i} A_{i} \frac{\partial F}{\partial x_{i}}+\frac{1}{2}\sum _{i,j} B_{ij}B_{ij}^{T} \frac{\partial^{2}F}{\partial x_{i}\partial x_{j}},
\end{eqnarray}
\normalsize
and time-forward:
\small
\begin{eqnarray}
\label{forwardFPE}
\frac{\partial F}{\partial t}=\sum _{i}\frac{\partial}{\partial x_{i}}(A_{i}\cdot F)+\frac{1}{2}\sum _{i,j}\frac{\partial^{2}}{\partial x_{i}\partial x_{j}}(B_{ij}B_{ij}^{T}\cdot F).
\end{eqnarray}
\normalsize
SDEs equivalent to  Eqs.~\ref{backwardFPE} and \ref{forwardFPE} has a form (e.g. \cite{Ga}):
\begin{eqnarray}\label{SDE1}
d\vec{r}=\vec{A_{i}}\cdot dt+B_{ij}\cdot d\vec{W},
\end{eqnarray}
where $\vec{r}$ is the individual pseudoparticle trajectory in the phase space and $dW_{i}$ is the Wiener process, commonly written as $dW_{i}=\sqrt{dt}\cdot dw_{i}$, with $dw_{i}$ being the randomly fluctuating term with Gaussian distribution. \\
To solve Eq. \ref{SDE1} in both cases (backward and forward), at the onset we initiate pseudoparticle at some starting point in space and time and integrate alongside the pseudoparticle trajectory until it reaches the boundary. 
Choosing the forward or backward approach we need to bear in mind the problem that we want to solve. In the forward integration, pseudoparticles start from diverse boundary points, being for the GCR particles the entrance to the heliosphere. After that, their trajectories are traced up to the target position, e.g. 1 AU. Thus, a high number of pseudoparticles has to be initialized in order of obtaining a robust statistic because plenty of them do not reach the target position. The backward integration is much more effective in the case of the GCR propagation in the heliosphere. In the backward approach, the number of 'useless' particles is reduced. Pseudoparticles start from point of interest (e.g.~1 AU) and are traced backward in time until crossing the heliosphere boundary (in this paper this boundary is assumed at 100 AU, Fig.~\ref{fig:fig1}). The value of the particle distribution function $f(\vec{r}, R)$ for the starting point can be found as an average of $f_{LIS}(R)$ value for  pseudoparticles characteristics at the entry positions, $f(\vec{r}, R)=\frac{1}{N}\sum_{n=1}^{N}f_{LIS}(R)$,
where $f_{LIS}(R)$ is the cosmic ray local interstellar spectrum taken as in \cite{Webber} for rigidity $R$ of the $n^{th}$ particle at the exit/entrance point.\\
The Parker equation (Eq. \ref{ParkerEq}) in the 3-D spherical coordinate system $(r,\theta,\varphi)$  as time-backward FPE diffusion equation has a form:
\small
\begin{eqnarray}
\label{backwardParker}
\frac{\partial f}{\partial t}=A_{1}\frac{\partial^{2} f}{\partial r^{2}}+A_{2}\frac{\partial^{2} f}{\partial \theta^{2}}+A_{3}\frac{\partial^{2} f}{\partial \varphi^{2}}+A_{4}\frac{\partial^{2} f}{\partial r \partial \theta}+A_{5}\frac{\partial^{2} f}{\partial r \partial \varphi}+A_{6}\frac{\partial^{2} f}{\partial \theta \partial \varphi}+A_{7}\frac{\partial f}{\partial r}+A_{8}\frac{\partial f}{\partial \theta}+A_{9}\frac{\partial f}{\partial \varphi}+A_{10}\frac{\partial f}{\partial R}
\end{eqnarray}
\normalsize
where coefficients have a form:\\
$A_{1}=K_{rr}^{S}, A_{2}=\frac{K_{\theta \theta}^{S}}{r^{2}}, A_{3}=\frac{K_{\varphi \varphi}^{S}}{r^{2}sin^{2} \theta},  A_{4}=\frac{2K_{r \theta }^{S}}{r}, A_{5}=\frac{2K_{r \varphi}^{S}}{r sin \theta}, A_{6}=\frac{2K_{\theta \varphi}^{S}}{r^{2}sin \theta}$, \\
$A_{7}=\frac{2}{r}K_{rr}^{S}+\frac{\partial K_{rr}^{S}}{\partial r}+\frac{ctg\theta}{r}K_{\theta r}^{S}+\frac{1}{r}\frac{\partial K_{\theta r}^{S}}{\partial \theta}+\frac{1}{r sin \theta}\frac{\partial K_{\varphi r}^{S}}{\partial \varphi}-U-v_{d,r}$,\\
$A_{8}=\frac{K_{r \theta}^{S}}{r^{2}}+\frac{1}{r}\frac{\partial K_{r \theta}^{S}}{\partial r}+\frac{1}{r^{2}}\frac{\partial K_{\theta \theta}^{S}}{\partial \theta}+\frac{ctg \theta}{r^{2}}K_{\theta \theta}^{S}+\frac{1}{r^{2} sin \theta}\frac{\partial K_{\varphi \theta}^{S}}{\partial \varphi}-\frac{1}{r}v_{d,\theta}$, \\
$A_{9}=\frac{K_{r \varphi}^{S}}{r^{2} sin\theta}+\frac{1}{r sin \theta}\frac{\partial K_{r \varphi}^{S}}{\partial r}+\frac{1}{r^{2}sin \theta}\frac{\partial K_{\theta \varphi}^{S}}{\partial \theta}+\frac{1}{r^{2} sin^{2} \theta}\frac{\partial K_{\varphi \varphi}^{S}}{\partial \varphi}-\frac{1}{r sin \theta}v_{d,\varphi}$,
$A_{10}=\frac{R}{3}\nabla \cdot U$.\\
\\
In our model we are using a full 3D anisotropic diffusion tensor of the GCR  particles $K_{ij}=K_{ij} ^{(S)}+K_{ij} ^{(A)}$ containing the symmetric $K_{ij} ^{(S)}$  and antisymmetric $K_{ij} ^{(A)}$  parts presented in \cite{Alania02}.  The drift velocity of the GCR particles  is realized as: $ v_{d,i}=\frac{\partial K_{ij} ^{(A)}}{\partial x_{j}}$ \cite{jokipii77}. The equivalent to Eq.~\ref{backwardParker} set of SDEs with matrix $B_{ij}$, $(i,j=r,\theta,\varphi)$ has a following form (the same can be found in \cite{Kopp2012}):

\begin{eqnarray}\label{SDE}
dr &=& A_{7} \cdot dt+[B \cdot dW]_{r} \nonumber \\
d \theta &=& A_{8}\cdot dt+[B \cdot dW]_{\theta}\\
d\varphi &=& A_{9}\cdot dt+[B\cdot dW]_{\varphi} \nonumber\\
dR &=& A_{10}\cdot dt,\nonumber
\end{eqnarray}

\[ B_{ij} = \left[ \begin{array}{ccc}
 \sqrt{2A_{1}} & 0 & 0 \\
\frac{A_{4}}{\sqrt{2A_{1}}} &  \sqrt{2A_{2}-\frac{A_{4}^{2}}{2A_{1}}} & 0 \\
\frac{A_{5}}{\sqrt{2A_{1}}} & \frac{A_{6}-\frac{A_{4}A_{5}}{2A_{1}}}{B_{\theta \theta}} & \sqrt{2A_{3}-B_{\varphi r}^{2}-B_{\varphi \theta}^{2}} \end{array} \right].\]

\begin{figure}[tbp]
  \begin{center}
\includegraphics[width=0.45\hsize]{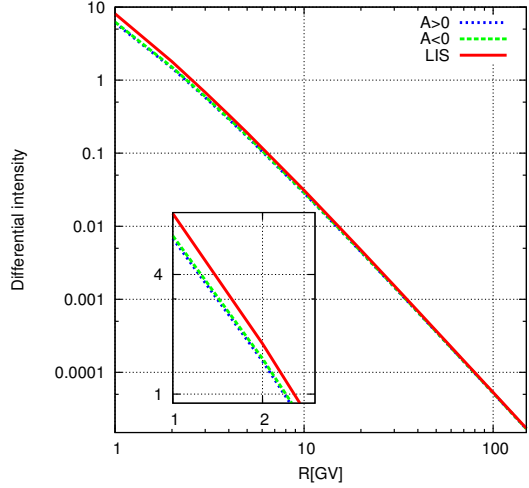}
\end{center}
\caption{\label{fig:spectrum} Simulated galactic protons rigidity spectra at Earth with respect to the LIS (solid line) at 100 AU for two cases: the $A > 0$ (dotted lines) and the $A < 0$ (dashed line) polarity epochs.}
\end{figure}

The Euler$-$Maruyama scheme is used in order to integrate Eqs. \ref{SDE} backward in time.
To solve Eqs. \ref{SDE} in spherical coordinates system we are using boundary conditions having form: $\varphi_{i}<0 \rightarrow \varphi_{i}=\varphi_{i}+2 \pi $, $\varphi_{i}>2 \pi \rightarrow \varphi_{i}=\varphi_{i}-2\pi $, $\theta_{i}<0 \rightarrow \theta_{i}=\theta_{i}+2\pi $ and  $\theta_{i}> \pi \rightarrow \theta_{i}=\pi - |\theta_{i}| $. The reflecting boundary is considered as the inner radial boundary, $\frac{\partial f}{\partial r}=0$ at $r=0.001$ AU. An empty heliosphere constitutes an initial condition: $f_{i}(0.01AU<r<100 AU, \theta,\varphi,R,0)=0$, as was shown in \cite{Pei}. Performed tests proved that the 3000 pseudoparticles were enough number to simulate the short time variations of GCR particles with rigidity 10 GV.\\
Fig.~\ref{fig:fig2} shows the trajectories of simulated pseudoparticles (galactic protons) for the positive ($A>0$) (upper panels) and for the negative ($A<0$) polarity epoch (bottom panels), initially injected with rigidity 10 GV from position $r=1AU$, $\theta=90^{\circ}$, $\varphi=180^{\circ}$.
The left panels in Fig.~\ref{fig:fig2} present the radial position of pseudoparticles versus backward time, while the right panels show the position in latitude versus time.
Fig.~\ref{fig:fig3} illustrates the binned exit rigidity of simulated pseudoparticles and corresponding binned propagation times for pseudoparticles. The exit time is a bit shorter for the $A>0$ than for the $A<0$ because of the large drift speed in the polar regions in $A>0$. This is in agreement with \cite{Zhang1999}.
Fig.~\ref{fig:scatter} displays latitude vs. longitude distribution of simulated pseudoparticles initiated at Earth orbit with rigidity 10 GV for $A>0$ and $A<0$ cycles. Fig.~\ref{fig:scatter} shows the different character of heliospheric transport depending on various drift patterns in the $A>0$ and $A<0$ cycles. In the $A>0$ cycle, pseudoparticles  are transported mainly toward higher latitudes, while in the $A<0$ period pseudoparticles are drifting outward mainly along the neutral sheet at low latitudes (Fig.~\ref{fig:fig2}, right panel).\\
Fig.~\ref{fig:spectrum} demonstrates simulated rigidity spectra at the Earth for the $A>0$ (dotted line), and for the $A<0$ (dashed line) polarity epoch. The solid line represents the unmodulated spectrum (LIS)  \cite{Webber} at 100 AU. However, for the rigidity greater than 1 GV difference is subtle, but still visible, being in an agreement with Zhang results (compare with \cite{Zhang1999}, Fig. 3).

\section{Model of the Forbush decrease of the GCR intensity}

\begin{figure}[t]
  \begin{center}
\includegraphics[width=0.5\hsize]{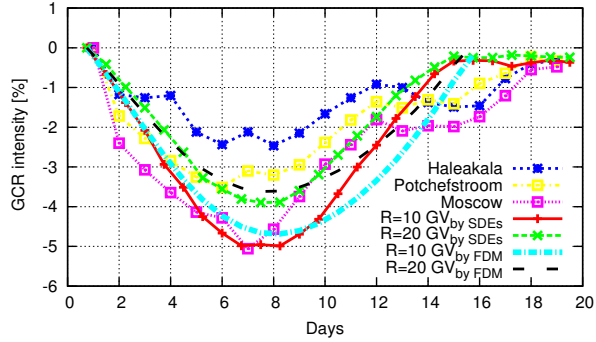}
\end{center}
\caption{\label{fig:ModelExper}Changes of the expected amplitudes of the Fd of the GCR intensity at the Earth orbit, for the rigidity of  10 and 20 GV based on the solutions of the Parker transport equation by SDEs and FDM in comparison with the GCR intensity registered by  Haleakala, Potchefstroom and Moscow neutron monitors during the Fd in 18 March - 4 April 2002.}

\end{figure}

We present the model of the recurrent Fd taking place due to established corotating heliolongitudinal disturbances in the interplanetary space. Corotating interaction regions (CIR) passing the Earth gradually diminishes the diffusion at the Earth orbit, causing larger scattering of the GCR particles, and in effect fewer GCR particles reach the Earth. We simulate this process by the gradual decrease and then the increase of the diffusion coefficient at the Earth orbit with respect the heliolongitudes. The diffusion coefficient $K_{\parallel}$ of cosmic ray particles has a form: $K_{\parallel}=K_{0}\cdot K(r)\cdot K(R,\nu)$, where $K_{0}=10^{21}cm^{2}/s$, $K(r)=1+0.5\cdot (r/1 AU)$ and $K(R,\nu)=R^{2-\nu}$. The exponent $\nu$ pronounces the increase of the HMF turbulence in the vicinity of space where the Fd is created (e.g. \cite{WA08, WA10}), and is taken as: $\nu = 1+ 0.3 sin(\varphi-90^{\circ})$ for $r < 30 AU$ and $90^{\circ}\leq \varphi \leq 270^{\circ}$. We assume the existence of the two dimensional spiral Parker's heliospheric magnetic field $B$ \cite{Parker58} implemented through the angle $\psi=arctan(-B_{\varphi}/B_{r})$ in the 3D anisotropic diffusion tensor $K_{ij}$ of GCR particles \cite{Alania02}. We compare the results obtained by the solution of the Parker transport equation by SDEs with our previous method of solution by FDM \cite{ WA10} assuming the same changes of all included parameters.\\
The expected changes of the GCR intensity for the rigidity of 10 and 20 GV during the simulated Fd in comparison with the profiles of the daily GCR intensities recorded by the three neutron monitors with different cut off rigidities in 18 March - 4 April 2002 presents Fig.~\ref{fig:ModelExper}.  One can see that the proposed models are in a good coincidence with the experimental data and, as is expected the amplitude of the Fd decreases for higher rigidities. Moreover, the model of the Fd obtained based on the solution of the SDEs allows to reflect the stochastic character of the GCR particles  distribution in the heliosphere and analyze the pseudoparticle trajectory thorough the 3D heliosphere (Fig.~\ref{fig:fig1}), which is not possible based on the solution of the Parker transport equation by the FDM (e.g.  \cite{WA08}). This feature can allow a thorough analysis of the trajectory of the GCR particles when encountering occasional shock waves accompanying the sporadic Fd.
\begin{figure}[t]
  \begin{center}
 \includegraphics[width=0.5\linewidth]{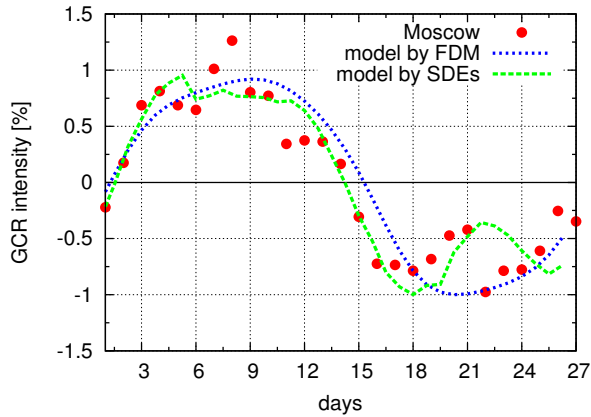}
\end{center}
 \caption{ \label{fig:ModelExper27}Changes of the expected amplitudes of the 27-day variation of the GCR intensity at the Earth orbit for the rigidity of  10 GV based on the solutions of the Parker transport equation by SDEs and FDM in comparison with the GCR intensity registered by Moscow neutron monitor for the 7 September- 3 October 2007.}
 \end{figure}

\section{Model of the 27-day variation of the GCR intensity}
The recurrence of the GCR intensity connected with the solar rotation is commonly called  the 27-day variation, although the durations can slightly differ. The 27-day variation of the GCR intensity is connected with the heliolongitudinal asymmetry of the electromagnetic conditions in the heliosphere. The recent minimum of solar activity between solar cycles No. 23 and 24 was quite exceptional. Recurrent variations connected with corotating structures ($\sim27$ days), at the end of 2007 and almost for the whole year 2008  were clearly established in all solar wind and interplanetary parameters. Consequently, the 27-day variations of cosmic ray intensity were clearly  visible in a variety of cosmic ray counts of neutron monitors (e.g., \cite{Alania10,Modz}) and space probes (e.g., \cite{Leske13}). We present the model of the 27-day variation of the GCR intensity considering an individual period of solar rotation starting at 2007.09.07. As it was stated by \cite{Alania10}-\cite{Gil13} the 27-day variation of the GCR intensity in the minimum epochs is preferentially related to the heliolongitudinal asymmetry of the solar wind velocity.
\indent In the model we apply approximation of the in situ measurements of the solar wind speed as source of the 27-day variation of the GCR intensity described by the formula:
$U =U_{0}(1-0.31sin(\varphi+6.10)+0.06sin(2\varphi+0.82)-0.10sin(3\varphi-1.04))$, $U_{0}=400 km/s$. \\
The diffusion coefficient $K_{\parallel}$ of cosmic ray particles has a form: $K_{\parallel}=K_{0}\cdot K(r)\cdot K(R)$, where $K_{0}=10^{22}cm^{2}/s$, $K(r)=1+0.5\cdot (r/1 AU)$ and $K(R)=R^{0.5} $.
 Fig. ~\ref{fig:ModelExper27} compares the results of solution of the Parker transport equation by SDEs with solution by FDM for the GCR particles with rigidity R=10 GV. The model of the 27-day wave of the GCR intensity obtained by the by SDEs and FDM at the Earth orbit (1 AU, $\theta=90^{0}$) is in agreement with the data of Moscow NM (Fig.~\ref{fig:ModelExper27}).

\section{Conclusion}
\begin{itemize}
  \item We presented the model of the Fd and the 27-day variation of the GCR intensity obtained based on the stochastic approach to the solution of the Parker transport equation. The modeling results are in a good agreement with the neutron monitors data.
  \item The SDEs were integrated backward in time in the spherical coordinates applying the full 3D anisotropic diffusion tensor.
  \item We showed the excellent agreement between the stochastic results and the finite difference method results presented in our previous papers.
\item The models obtained based on the solution of the SDEs allow to reflect the stochastic character of the GCR particles  distribution in the heliosphere. Additionally this approach allows to trace the pseudoparticle trajectory through the 3D heliosphere which is not possible based on the solution of the Parker transport equation by finite difference methods.

\end{itemize}

\section*{Acknowledgments}
This work is supported by The Polish National Science Centre grant awarded by decision number DEC-2012/07/D/ST6/02488. We thank the principal investigators of Potchefstroom, Moscow and Haleakala neutron monitors for the ability to use their data.

\end{document}